# ITO-Based Microheaters for Reversible Multi-Stage Switching of Phase-Change Materials: Towards Miniaturized Beyond-Binary Reconfigurable Integrated Photonics

Hossein Taghinejad,[1, Δ] Sajjad Abdollahramezani,[1, Δ] Ali A. Eftekhar,[1, Δ] Tianren Fan,[1] Amir H. Hosseinnia,[1] Omid Hemmatyar,[1] Ali Eshaghian Dorche,[1] Alexander Gallmon,[2] and Ali Adibi [1, *]

1: School of Electrical and Computer Engineering, Georgia Institute of Technology, Atlanta, GA 30332, USA. 2: Institute of Electronics and Nanotechnology, Georgia Institute of Technology, Atlanta, GA 30332, USA. Δ: Authors with equal contributions.

*: ali.adibi@ece.gatech.edu

*Abstract.* Inducing a large refractive-index change is the holy grail of reconfigurable photonic structures, a goal that has long been the driving force behind the discovery of new optical material platforms. Recently, the unprecedentedly large refractive-index contrast between the amorphous and crystalline states of Ge-Sb-Te (GST)-based phase-change materials (PCMs) has attracted a tremendous attention for reconfigurable integrated nanophotonics. Here, we introduce a microheater platform that employs optically transparent and electrically conductive indium-tin-oxide (ITO) bridges for the fast and reversible electrical switching of the GST phase between crystalline and amorphous states. By the proper assignment of electrical pulses applied to the ITO microheater, we show that our platform allows for the registration of virtually any intermediate crystalline state into the GST film integrated on the top of the designed microheaters. More importantly, we demonstrate the full reversibility of the GST phase between amorphous and crystalline states. To show the feasibility of using this hybrid GST/ITO platform for miniaturized integrated nanophotonic structures, we integrate our designed microheaters into the arms of a Mach-Zehnder interferometer to realize electrically reconfigurable optical phase shifters with orders of magnitude smaller footprints compared to existing integrated photonic architectures. We show that the phase of optical signals can be gradually shifted in multiple intermediate states using a structure that can potentially be smaller than a single wavelength. We believe that our study showcases the possibility of forming a whole new class of miniaturized reconfigurable integrated nanophotonics using beyond-binary reconfiguration of optical functionalities in hybrid PCM-photonic devices.

**Introduction**

Manipulation of the refractive index forms the backbone of reconfigurable integrated photonics, a notion that constantly seeks alternative material platforms and modulation schemes for achieving new functionalities and better performance measures. [1–4] The latest trends show a growing interest in hybrid photonic devices based on phase-change materials (PCMs) with a specific focus on germanium-antimony-telluride (Ge-Sb-Te or shortly GST) alloys for operation in the infrared wavelengths. This increasing attention stems from the unprecedented large, reversible, and non-volatile change in the real (n) and imaginary (k) parts of GST refractive index upon the phase transition between amorphous and crystalline states. For instance, at the telecommunication wavelength of 1550 nm, GST provides an index change of $\Delta n > 2$, [5] which is orders of magnitude larger than what can be obtained via alternative methods such as the thermo-optic effect in silicon with $\Delta n /\Delta T \sim 1.86\times10^{-4}$ K$^{-1}$, [6] corresponding to $\Delta n \approx 0.054$ at $\Delta T = 300$ °C. Access to such a large $\Delta n$ budget enables the design of phase shifters, modulators, and switches with ultra-compact footprints, potentially down to subwavelength scales. [7–11] In addition, the induced index-change is nonvolatile, fast (a few ns), and reversible for many overwriting cycles ($> 10^{15}$), a collection of features that ensures the low-power, high-speed, and long-term reconfiguration of GST-based photonic devices. [1,4,12,13] These features have made GST alloys very appealing for reconfigurable photonic platforms such as all-optical memories, [14,15] active meta-surfaces, [16–18] neuromorphic computing, [7,19,20] and optical switching. [10,21,22]

The success of reconfigurable PCM-based photonic platforms is entangled with the reliable control of the PCM phase, which can be reversibly switched between amorphous and crystalline states through a controlled heating process. Heating GST at a moderate temperature (150-250 °C) induces transition from amorphous to crystalline, and the melting of GST at temperatures above 600 °C followed by a fast cooling (>1 °C/ns) results in the reverse conversion from crystalline to amorphous, a process that is widely referred to as the melt quenching. [23–25] The required heat for phase switching can be supplied directly through heating or indirectly by optical and/or electrical stimuli. The use of direct heating, however, has been limited to preliminary proof-of-concept demonstrations as it only allows for the one-way amorphous-to-crystalline conversion. On the other hand, the reversible switching of the GST phase [26,27] as well as the selective conversion of individual GST inclusions have been demonstrated using the optical stimulation. [16,28] On the downside, however, optical stimulation necessitates bulky external laser sources or complex none-monolithic integration schemes, which hinder the implementation of fully integrated device platforms for practical applications. Moreover, because of the low optical absorption and small absorption cross-section of amorphous GST nanostructures, recrystallization cycles require high-power optical pulses. [29]

Among different options, electrical heating, also known as Joule heating, holds a great promise for the ultimate miniaturization of GST-based active and reconfigurable photonic devices, primarily because the miniaturized heaters can be integrated into the device platform. In its simplest form, the Joule heating can be implemented by flowing an electrical current (I) directly through the GST film. [30] In this scheme, however, the large variation of GST resistance ($R_{GST}$) between amorphous and crystalline states requires a large variation of input current to generate enough heat for the phase conversion (heat varies linearly with $R_{GST}I^2$). [31] This issue can be addressed by decoupling GST from the current path, potentially through the integration of GST on top of a microheater, so that the Joule heating can be controlled in the microheater (independent of the GST phase). Designing a microheater that confines the heating volume for efficient GST heating, provides a short thermal-time-constant, and does not add optical loss is challenging. The use of metals with large optical loss should be refrained in most photonic applications. As a solution, a recent study proposed the use of two-dimensional graphene for designing microheaters with low optical loss. [32] However, because of abundant defects and grain boundaries, graphene is prone to fast oxidation in the ambient, [33] which makes its long-term operation under recurring heat cycles unreliable. In addition, the integration of graphene with photonics devices requires a mechanical transfer procedure that is not compatible with mainstream fabrication processes. Other reported methods such as doped-silicon microheaters [34–36] are not holistic and cannot be adopted for alternative material systems such as silicon nitride (SiN) and silicon carbide (SiC) as promising material platforms for integrated nanophotonics. Thus, designing a compact, low-loss, and CMOS-compatible microheater is a major bottleneck for the practical deployment of integrated photonic solutions that utilize PCMs for reconfiguration.

Here, we report the realization of a microheater platform that employs optically transparent (i.e., low-loss) and electrically conductive indium-tin-oxide (ITO) bridges for the fast and reversible switching of the GST phase between crystalline and amorphous states. By the proper assignment of width and amplitude of electrical pulses applied to the ITO bridge, we show that our platform allows for the registration of virtually any intermediate crystalline state into a GST film integrated on the top of the designed microheater. Our estimation of energy-per-pulse consumed for the amorphous-to-crystalline phase conversion in GST is ~ 6.5 nJ in our test devices with the possibility of sub-nJ operation in an optimal device configuration. Granted by the low optical loss of the ITO films, we further show that our designed microheater platform can be directly integrated on SiN-based photonic devices for the realization of electrically reconfigurable optical functionalities. In a representative demonstration, we integrate our designed ITO-based microheaters into the arms of a Mach-Zehnder interferometer (MZI) to realize electrically reconfigurable optical phase shifters. By applying optimized electrical pulses to microheaters, the phase of optical signals can be gradually shifted between 0-50° with multiple intermediate values corresponding to the intermediate crystalline states of GST.

As schematically shown in Figures 1(a, b), our designed microheater relies on the Joule heating in a 60 nm-thick ITO bridge that is electrically driven via two gold/titanium (Au/Ti) electrodes formed on its two ends. ITO is deposited by the DC sputtering of an indium-oxide/tin-oxide ($In_2O_3$/$SnO_2$ with 90/10 wt %) target in an argon (Ar) plasma (~ 70 W, ~ $5 \times 10^{-3}$ Torr). Subsequently, electrodes are formed by the deposition of a 150/20 nm Au/Ti stack in an electron-beam (e-beam) evaporation chamber at the base pressure of ~ $6 \times 10^{-6}$ Torr. Finally, we integrate a 30 nm-thick GST patch on top of the ITO bridge via the RF sputtering of a stoichiometric $Ge_2Sb_2Te_5$ target. Optical and scanning-electron microscope (SEM) micrographs of finalized devices are shown in Figures 1(c) and 1(d), respectively. The lateral dimensions of the GST patch and, accordingly, the ITO bridge is intentionally designed to be large for convenient characterization of the designed platform. However, as we will show in the following sections, dimensions can be scaled down to fit the requirements of the miniaturized device fabrication. In addition, our sputtering conditions (i.e., RF power of 40 W and the vacuum level of $10^{-2}$ Torr) are optimized so that the stoichiometry of the deposited GST film matches that of the sputtering target. Under these conditions, we obtain a deposition rate of ~ 1.7 nm/min (Figure 1€) and a thickness uniformity of better than 10% across a two-inch wafer. Such a uniformity leads to a negligible thickness variation across our small device footprints (Figure 1(f)).

The electrical conductivity of the ITO bridge is an important design parameter that directly affects the efficiency of the Joule heating. The sputtering deposition of ITO typically yields a thin film with relatively small crystalline domains, leading to a suboptimal electrical conductivity in as-deposited films. However, the post-deposition annealing is shown to be effective for enhancing the electrical conductivity of ITO thin films, primarily via increasing crystalline-domain sizes and activating Sn donors. [37,38] Thus, we perform a rapid thermal annealing (RTA; 15 mins @ 450 °C) on our sputtered ITO, which reduces the sheet resistance of the film from ~3000 Ω/sq (before RTA) to ~100 Ω/sq (after RTA). Additionally, our RTA process is conducted under a mild flow of oxygen, which, according to a previous study, [39] reduces the optical loss of ITO thin films, a critical feature for the direct integration of our microheater with photonic devices without the addition of extra optical loss. Furthermore, the RTA process immunes the ITO bridge from random resistance variations during the heat cycles generated for the phase conversion of GST patches.

Before testing, we covered the surface of the fabricated device in Figure 1 with a 200 nm-thick silicon dioxide ($SiO_2$) capping layer. The role of the capping layer is threefold: (i) protecting the GST patch from gradual oxidation in the ambient (see Figure S1, Supporting Information (SI)), (ii) preventing the decomposition of GST during the heating cycles, and (iii) precluding the failure of the microheater caused by the electric breakdown of the air at the sharp corners of the device. Accordingly, with the $SiO_2$ capping layer, (i) samples can be kept in the ambient for an arbitrarily long period of time, (ii) the GST patch can be successfully maintained during the full cycle of the amorphous-to-crystalline conversion, and (iii) the applied voltage can be increased to generate sufficient heat for the phase conversion of the GST patch integrated on the microheater.

For optical characterization, we primarily rely on the change of refractive index (Δn) and extinction coefficient (Δk) of GST as we electrically derive a gradual transition from the amorphous to crystalline phase via Joule heating of the GST patch. Thus, we use optical reflection spectroscopy to demonstrate the electrical control of the GST phase from the amorphous to the full-crystalline state with multiple intermediate states. As shown in Figure 2(a), after each Joule heating event, the reflection of a broadband light source from the center of the GST patch was collected via a 20X objective lens (numerical aperture, NA ≈ 0.42) and normalized to that obtained from the ITO region without GST. The measured reflection spectra contain multiple peaks and dips, which are generated by the light interference inside the stack of layers constituting the device. The position and amplitude of these reflection features are dictated by the optical constants of various layers in the stack, among which only those of the GST layer alter upon the Joule heating events. Therefore, the modulation of peaks and dips in the reflection spectrum of a device can be used for the back-calculation of the optical properties of the GST layer and to estimate its crystalline phase after each Joule heating event.

To extract the degree of crystallization from the measured reflection spectra, we need to benchmark the estimated n and k values of GST against a reference dataset. Thus, following previous reports, [31] we vacuum-annealed an amorphous GST layer at ~ 250 °C for 5 mins, which provides a reference sample for the crystalline state of GST. Subsequently, we used ellipsometry measurements with a Tauc-Lorentz dispersion fitting [5] to directly obtain n and k of reference crystalline/amorphous GST films as depicted with solid lines in Figures 2(b) and 2(c), respectively. To obtain similar references for the intermediate states of GST, we used an effective-medium approximation in which an α fraction of the GST volume is assumed to be in the crystalline state and the (1-α) fraction of it in the amorphous state (0 < α < 1). Thus, using the measured n and k values of amorphous (i.e., α = 0) and crystalline (i.e., α = 1) GST films, we can extrapolate n and k values for any arbitrary intermediate GST state using the Lorentz-Lorenz equation: [40]

$$\varepsilon_{eff}(\alpha) = \frac{1+2T}{1-T}, \qquad T = \alpha \frac{\varepsilon_c - 1}{\varepsilon_c + 2} + (1-\alpha)\frac{\varepsilon_a - 1}{\varepsilon_a + 2} \qquad \text{Eq. (1)}$$

$$n_{eff}(\alpha) + ik_{eff}(\alpha) = \sqrt{\varepsilon_{eff}} \qquad \text{Eq. (2)}$$

In Eq. (1), the α-dependent dielectric constant of the effective medium ($\varepsilon_{eff}(\alpha)$) can be calculated from the dielectric constants of crystalline ($\varepsilon_c$) and amorphous ($\varepsilon_a$) GST states. Subsequently, using Eq. (2), the real ($n_{eff}$) and imaginary ($k_{eff}$) parts of the effective refractive index can be derived from the $\varepsilon_{eff}(\alpha)$. Some of the extrapolated n and k values for intermediate GST states with α ≠ 0, 1 are plotted with dashed lines in Figures 2(b) and (c).

To estimate the crystallization level after each Joule heating event, we employ the transfer-matrix method to regenerate the experimental reflection spectra shown in Figure 2(a). In these calculations, we model each material layer in Figure 1(a) by its thickness and optical (*n* and *k*) constants. We use $n_{eff}(\alpha)$ and $k_{eff}(\alpha)$ as the optical constants of the GST layer, and, following a brute-force approach, we sweep the α parameter from 0 to 1 and look for the best fit to the experimental results. Our calculations show that the reflection spectra displayed in Figure 2(a) can be successfully fitted with α = 0, 0.3 ± 0.1, 0.6 ± 0.1, and 0.9 ± 0.1. We also note that, the optical spectra in Figure 2(a) are measured with the microheater switched off after each heating step to evaluate the nonvolatile nature of the induced GST phase change. Thus, our results in Figure 2 clearly demonstrate the ability of our designed ITO microheater for nonvolatile electrical control of the GST phase via Joule heating.

The change of the GST phase can be further studied through monitoring its electrical properties, especially for miniaturized dimensions at which optical characterizations are complicated. As shown in Figure 3(a), we design and fabricate a four-electrode device in which two readout electrodes are placed across the GST patch for the electrical characterization of the GST phase following sequential Joule-heating pulses applied to the ITO microheater. To electrically isolate GST from the underlying ITO film, a 10 nm-thick hafnium dioxide ($HfO_2$) spacer layer is deposited (using atomic layer deposition, ALD) between the GST patch and the ITO microheater. For the amorphous-to-crystalline conversion, we apply 200 msec voltage pulses of varying amplitudes to the microheater, and after each pulse we measure the GST resistance ($R_{GST}$) using the readout electrodes. As shown in Figure 3(b), with increasing the height and the number of applied pulses, $R_{GST}$ monotonically drops with a staircase profile, meaning that the GST phase is gradually changing from the amorphous state (high-resistance) to the crystalline state (low-resistance). Indeed, each distinguishable $R_{GST}$ value represents a different GST (crystalline) phase, showing that GST can be configured in virtually any intermediate crystalline state through the controlled excitation of the ITO microheater. However, the impact of the pulse amplitude on changing the GST phase is more significant than the number of applied pulses. In fact, at a fixed pulse amplitude, the minimum achievable resistance saturates with the number of applied pulses. This observation suggests that the crystallization of GST is primarily a temperature-driven phenomenon and obtaining higher crystallization levels mandates the temperature increase via the application of an electrical pulse with a larger amplitude.

The GST phase can be switched back to the amorphous state by melt-quenching, a process that needs the high-temperature melting and fast cooling of crystalized GST. Thus, we increase the amplitude of the voltage pulse to ~9 V, for melting, and narrow the pulse width down to 50 nsec for the fast cooling of the GST patch. As indicated by an arrow in Figure 3(b), the application of this short pulse to the microheater successfully resets the GST resistance back to the amorphous level, which verifies the feasibility of reversible switching in our developed ITO-based microheater platform. Interestingly, the re-crystallization of the melt-quenched GST (i.e., the response in Figure 3(b) after pulse index 125) displays a sudden drop in $R_{GST}$, which is then followed by the recovery of the stepwise resistance drop in the subsequent crystallization steps. In other words, the number of achievable intermediate crystalline states in the melt-quenched GST is less than that in the as-deposited amorphous film. Previous studies suggest that such a difference stems from the residual crystalline domains in the melt-quenched GST film, which bypasses the nucleation step that is required for the crystallization of as-deposited GST film.[30] Therefore, compared to as-deposited films, an accelerated crystallization is anticipated in melt-quenched GST films. Nevertheless, our platform offers reversible switching of the GST phase with multiple intermediate crystalline states in a repeatable way. We repeated this experiment several times and noticed that the subsequent crystallization cycles (not shown here) follow consistently the trend observed in the second cycle, hinting that a pre-conditioning step is required prior to the long-term operation of the device. To the best of our knowledge, this is the first demonstration of an optically transparent microheater architecture for reversible multi-stage GST phase change with an electrical stimulus.

Figure 3(b) shows one order-of-magnitude change in $R_{GST}$ upon GST phase transition from amorphous to crystalline. This is noticeably smaller than the expected two-orders-of-magnitude change, as reported elsewhere.[11] We attribute this discrepancy to the heat-sinking effect of the large metallic (Au) electrodes (see Figure 3(a)) used for the readout of $R_{GST}$. The overlap of these Au electrodes with the ITO bridge disturbs the otherwise-uniform heat profile of the microheater and creates two locally cold regions close to the left and right metallic electrodes in Figure 3(a), which leaves behind two amorphous segments on the sides of the central crystalline GST segment (see Figure S2 in SI). In other words, the measured resistance (i.e., $R_{GST}$) in Figure 3(a) is composed of the series connection of three resistive components: two residual amorphous segments at the vicinity of the left and right electrodes (i.e., $R_l$ and $R_r$ in Figure 3(c)) and a middle segment (i.e., $R_m$) that can be switched between amorphous and crystalline states; that is, $R_{GST} = R_l + R_m + R_r$, as shown in Figure 3(c). Therefore, the smallest achievable resistance in the crystalline state is $R_{GST} = R_l + R_m + R_r \approx R_l + R_r$, (with $R_m$ being small in the crystalline phase), which limits the attainable resistance contrast. These residual amorphous segments, however, can be crystallized by the direct flow of electrical current through the Au electrodes for the local Joule heating at the vicinity of the readout electrodes. As shown in the shaded region of Figure 3(c), following the sequential application of current pulses (200 µA, 200 msec) to the Au electrodes, $R_{GST}$ can be further reduced by an additional order-of-magnitude to achieve the overall two-orders-of-magnitude expected resistance contrast. In practical photonic and optoelectronic devices, there is no need for the measurement of $R_{GST}$ through such large Au electrodes, and thus, this issue does not put any limitation on the capability of the demonstrated microheater architecture for practical applications.

We further study the spatial and temporal temperature profiles of the device in Figure 1(a) in response to the voltage pulses applied to the ITO microheaters by solving electro-thermal equations in the COMSOL Multiphysics simulation package for the experimental condition explained in Figure 3 (see SI for details). As shown in Figure 4(a), despite some expected temperature drop across the ITO width, the temperature profile is spatially uniform along the ITO bridge, suggesting that a GST patch located at the center of the microheater can be uniformly switched between amorphous and crystalline states. In addition to the spatial temperature uniformity, our designed platform ensures the fast thermal dynamics needed for the re-amorphization of crystalline GST. As shown in Figure 4(b), a cooling rate of ~ 1.3 °C/nsec is achievable following the trailing edge of a 50 nsec-wide 9 V pulse, close to the reported value of 1 °C/nsec for the melt-quenching of crystalline GST. A similar thermal transient simulation for the amorphous-to-crystalline phase conversion is also depicted in Figure 4 (c), showing that a 3V voltage pulse with a 300 nsec pulsewidth can elevate the GST temperature to ~ 350 °C for the full crystallization of an amorphous film.

Combining the experimental results (Figure 3) and electro-thermal simulations (Figure 4), we can estimate the performance measures the platform in Figure 3(a). As tabulated in Figure 4 (d), the energy-per-pulse needed for the crystallization of the amorphous GST and the amorphization of the crystalline GST in Figure 3(a) are ~ 6.6 nJ and 10 nJ, respectively. However, we note that these values are overestimated, and the real energy consumption in an actual device (i.e., not a test device) is noticeably smaller because of two main reasons. First, the readout electrodes, used in the test device, sink a portion of the generated heat that could be otherwise consumed for the GST phase conversion. Second, the $HfO_2$ layer, used for the electrical isolation of the ITO bridge and readout electrodes (for enabling the measurement of $R_{GST}$) adds a thermal barrier between ITO and GST. Thus, the generated heat in the ITO microheater may not be fully delivered to the GST patch during short voltage pulses employed for the phase conversion. Elimination of these two factors can potentially lead to sub-nJ energy consumption in our hybrid ITO/GST platform.

To show the feasibility of forming miniaturized reconfigurable integrated photonic devices using our developed platform, we integrate the hybrid GST/ITO structure on top of a SiN waveguide, where the phase and amplitude of the optical transmission can be modulated by adjusting the GST crystalline state between the fully amorphous and fully crystalline phases. The SiN waveguide here is fabricated in a 400 nm-thick SiN film deposited by the low-pressure chemical vapor deposition (LPCVD) method. The waveguide width is set to ~ 1400 nm to ensure the single-mode operation around the wavelength of 1550 nm. The geometry of the GST nanostructure is tailored to optimize the ratio of the optical phase-shift to the transmission loss upon the full conversion from the amorphous to the crystalline phase of GST, while maintaining a very small footprint. The geometry of the ITO microheater is also designed to grant a uniform temperature profile across the GST patch, rapid thermal reconfiguration, and minimum optical loss. The electrical connections to the ITO bridge are supported by two small SiN sidebars (width ≈ 500 nm), which enables a low-resistance electrical connection to the ITO microheater with minimal extra waveguide loss (< 0.2 dB). The thickness of the ITO and GST layers are 60 nm and 30 nm, respectively.

To characterize the waveguide phase and amplitude changes upon the GST conversion, we integrate the designed phase shifter into the arm of a MZI structure. Figures 5 (a) and (b) show the optical image of the fabricated MZI and the false-colored SEM image of the GST/ITO microheater integrated on top of the SiN waveguide segment. To characterize the device, we measure the transmission spectrum of the MZI at various crystalline states of the GST patch. The GST conversion is induced by exciting the ITO heater using a train of voltage pulses. In this approach, the crystalline state of the GST can be controlled by adjusting the amplitude, width, or the number of the applied pluses as discussed before. Figure 5(c) shows the transmission spectrum of the MZI under the application of voltage pulses to the ITO microheater. The spectral shift and increased asymmetry in the spectral features of the MZI response are indicative of the amplitude modulation and phase shift in the waveguide induced by the nonvolatile change of the GST crystalline state. Figure 5(d) shows the extracted phase shift and amplitude of the integrated waveguide versus the applied voltage. The GST patch provides a reconfigurable phase shift from 0 to 50 degrees; however, the phase shift is accompanied with a relatively large transmission loss of ~ 7 dB.

We believe that the extra loss of the device is primarily due to the material loss of the GST in the crystalline phase. Considering the existing reports for GST with considerably lower values of $k$ at the 1550 nm wavelength, we think that further optimization of our deposition process can reduce the overall loss of the device. In addition, at the cost of making the structure a little larger, the loss can be reduced via mode engineering in the hybrid SiN/ITO/GST waveguide to shift a portion of the mode energy from GST to SiN upon conversion from the amorphous to the crystalline phase. Finally, the use of alternative GST-like alloys with lower $k$ values (e.g., $Ge_2Sb_2Se_4Te_1$) [2] can further reduce the loss at the expense of relatively lower operation speeds. We note that our developed microheater can be used for almost all PCMs, and the change of material will not cause any major change in our proposed fabrication process.

A key advantage of our phase-shifter structure in Figure 5 is that it achieves the phase shift of 50 degrees using a SiN waveguide segment with a length < 2 μm. To achieve the same phase shift using a conventional Si waveguide with the free-carrier-plasma-dispersion effect at the 1550 nm wavelength using similar voltages to our structure, a waveguide length of ~ 1mm is needed. Achieving the same phase shift with similar voltages in lithium niobate ($LiNbO_3$) requires a waveguide length of ~ 2 mm. Thus, our platform offers 3 orders-of-magnitude reduction in the device size at the same operation voltage. This can revolutionize the field of reconfigurable integrated nanophotonics for applications where reconfiguration times of 10's of ns are sufficient. We note that phase shifters using the thermo-optic effect in Si can have one order-of-magnitude smaller size than other conventional mechanisms (e.g., ~ 0.2 mm in this case). However, their excessive power consumption (due to the volatility of the mechanism and the need for constant heating) and slow response hinder their practical use. While the demonstrated results in Figure 5 only show the proof-of-principle for the feasibility of such integrated photonic devices, this unique advantage combined with the nonvolatility (and thus, potential ultra-low-power operation) can enable a reliable platform for a large range of reconfigurable devices. We believe through optimizing the waveguide device architecture and the material deposition and fabrication processes, our hybrid material platform with the electrically stimulated microheater architecture can serve as a reliable platform for miniaturizing reconfigurable integrated photonic structures.

To obtain a better understanding of the field profile of the structure in Figure 5, we use finite-difference time-domain (FDTD) analysis in the Lumerical simulation package to study the overlap of the electric field of the SiN waveguide mode with the GST region in its different crystalline phases (Figure 5(e)). The optical constants of GST and ITO are extracted from ellipsometry measurements. Our calculations show that the integration of our designed microheater platform with the SiN waveguide results in a relatively low loss of ~ 0.2 dB at the target operation wavelength (1550-1600 mn). Additionally, the as-deposited GST film (which is in the amorphous phase) contributes an additional loss of less than 0.6 dB, which is due to a combination of the optical absorption in the amorphous GST film and the scattering loss caused by the large refractive index of GST as compared to that of SiN. As Figure 5(e) represents, following the GST crystallization, the overlap of the electric field with the GST segment increases, which leads to the modulation of the phase and amplitude of the optical signal inside the SiN waveguide. Such an enhanced field overlap and phase/amplitude modulation stem from the larger refractive index of the crystalline GST as compared to the amorphous GST. We also note that the small refractive index contrast between SiN and ITO extends the electric field from the SiN waveguide into the ITO layer, which enhances the field interaction with the GST film on top of the ITO layer.

In conclusion, we introduced here an ITO-based microheater platform for the fast and reversible switching of the GST phase between crystalline and amorphous states. Our experimental demonstration and thermal analysis confirmed that controlling the Joule heating in the ITO heater, by the application of proper electrical pulses, enables recording virtually any intermediate crystalline state into GST films in a reversible way. We further employed this unique feature for the proof-of-concept demonstration of optical phase-shifting via the integration of a GST/ITO microheater on top of a SiN waveguide. We believe that our developed platform can be universally adopted for precise and reversible phase switching in other volatile and nonvolatile PCM materials for realization reconfigurable optoelectronic devices with miniaturized footprints.

**Acknowledgement**

This work was funded by Defense Advanced Research Projects Agency (DARPA) (GR10003430, Dr. Gordon A. Keeler). This work was performed in part at the Georgia Tech Institute for Electronics and Nanotechnology (IEN), a member of the National Nanotechnology Coordinated Infrastructure, which is supported by the National Science Foundation under Grant No. ECCS-1542174.

# References

(1) Wuttig, M.; Bhaskaran, H.; Taubner, T. Phase-Change Materials for Non-Volatile Photonic Applications. *Nat. Photonics* **2017**, *11* (8), 465–476. https://doi.org/10.1038/nphoton.2017.126.

(2) Zhang, Y.; Chou, J. B.; Li, J.; Li, H.; Du, Q.; Yadav, A.; Zhou, S.; Shalaginov, M. Y.; Fang, Z.; Zhong, H.; Roberts, C.; Robinson, P.; Bohlin, B.; Ríos, C.; Lin, H.; Kang, M.; Gu, T.; Warner, J.; Liberman, V.; Richardson, K.; Hu, J. Broadband Transparent Optical Phase Change Materials for High-Performance Nonvolatile Photonics. *Nat. Commun.* **2019**, *10* (1), 1–9. https://doi.org/10.1038/s41467-019-12196-4.

(3) Taghinejad, M.; Cai, W. All-Optical Control of Light in Micro- and Nanophotonics. *ACS Photonics* **2019**, *6* (5), 1082–1093. https://doi.org/10.1021/acsphotonics.9b00013.

(4) Abdollahramezani, S.; Hemmatyar, O.; Taghinejad, H.; Krasnok, A.; Kiarashinejad, Y.; Zandehshahvar, M.; Alu, A.; Adibi, A. Tunable Nanophotonics Enabled by Chalcogenide Phase-Change Materials. *ArXiv200106335 Phys.* **2020**.

(5) Shportko, K.; Kremers, S.; Woda, M.; Lencer, D.; Robertson, J.; Wuttig, M. Resonant Bonding in Crystalline Phase-Change Materials. *Nat. Mater.* **2008**, *7* (8), 653–658. https://doi.org/10.1038/nmat2226.

(6) Guha, B.; Gondarenko, A.; Lipson, M. Minimizing Temperature Sensitivity of Silicon Mach-Zehnder Interferometers. *Opt. Express* **2010**, *18* (3), 1879–1887. https://doi.org/10.1364/OE.18.001879.

(7) Feldmann, J.; Youngblood, N.; Wright, C. D.; Bhaskaran, H.; Pernice, W. H. P. All-Optical Spiking Neurosynaptic Networks with Self-Learning Capabilities. *Nature* **2019**, *569* (7755), 208–214. https://doi.org/10.1038/s41586-019-1157-8.

(8) Wu, C.; Yu, H.; Li, H.; Zhang, X.; Takeuchi, I.; Li, M. Low-Loss Integrated Photonic Switch Using Subwavelength Patterned Phase Change Material. *ACS Photonics* **2019**, *6* (1), 87–92. https://doi.org/10.1021/acsphotonics.8b01516.

(9) Stegmaier, M.; Ríos, C.; Bhaskaran, H.; Wright, C. D.; Pernice, W. H. P. Nonvolatile All-Optical 1 × 2 Switch for Chipscale Photonic Networks. *Adv. Opt. Mater.* **2017**, *5* (1), 1600346. https://doi.org/10.1002/adom.201600346.

(10) Zheng, J.; Fang, Z.; Wu, C.; Zhu, S.; Xu, P.; Doylend, J. K.; Deshmukh, S.; Pop, E.; Dunham, S.; Li, M.; Majumdar, A. Nonvolatile Electrically Reconfigurable Integrated Photonic Switch. *ArXiv191207680 Phys.* **2019**.

(11) Farmakidis, N.; Youngblood, N.; Li, X.; Tan, J.; Swett, J. L.; Cheng, Z.; Wright, C. D.; Pernice, W. H. P.; Bhaskaran, H. Plasmonic Nanogap Enhanced Phase-Change Devices with Dual Electrical-Optical Functionality. *Sci. Adv.* **2019**, *5* (11), eaaw2687. https://doi.org/10.1126/sciadv.aaw2687.

(12) Ding, F.; Yang, Y.; Bozhevolnyi, S. I. Dynamic Metasurfaces Using Phase-Change Chalcogenides. *Adv. Opt. Mater.* **2019**, *7* (14), 1801709. https://doi.org/10.1002/adom.201801709.

(13) Cao, T.; Cen, M. Fundamentals and Applications of Chalcogenide Phase-Change Material Photonics. *Adv. Theory Simul.* **2019**, *2* (8), 1900094. https://doi.org/10.1002/adts.201900094.

(14) Ríos, C.; Stegmaier, M.; Hosseini, P.; Wang, D.; Scherer, T.; Wright, C. D.; Bhaskaran, H.; Pernice, W. H. P. Integrated All-Photonic Non-Volatile Multi-Level Memory. *Nat. Photonics* **2015**, *9* (11), 725–732. https://doi.org/10.1038/nphoton.2015.182.

(15) Rios, C.; Hosseini, P.; Wright, C. D.; Bhaskaran, H.; Pernice, W. H. P. On-Chip Photonic Memory Elements Employing Phase-Change Materials. *Adv. Mater.* **2014**, *26* (9), 1372–1377. https://doi.org/10.1002/adma.201304476.

(16) Wang, Q.; Rogers, E. T. F.; Gholipour, B.; Wang, C.-M.; Yuan, G.; Teng, J.; Zheludev, N. I. Optically Reconfigurable Metasurfaces and Photonic Devices Based on Phase Change Materials. *Nat. Photonics* **2016**, *10* (1), 60–65. https://doi.org/10.1038/nphoton.2015.247.

(17) Yin, X.; Steinle, T.; Huang, L.; Taubner, T.; Wuttig, M.; Zentgraf, T.; Giessen, H. Beam Switching and Bifocal Zoom Lensing Using Active Plasmonic Metasurfaces. *Light Sci. Appl.* **2017**, *6* (7), e17016–e17016. https://doi.org/10.1038/lsa.2017.16.

(18) Abdollahramezani, S.; Taghinejad, H.; Fan, T.; Kiarashinejad, Y.; Eftekhar, A. A.; Adibi, A. Reconfigurable Multifunctional Metasurfaces Employing Hybrid Phase-Change Plasmonic Architecture. *ArXiv180908907 Phys.* **2018**.

(19) Feldmann, J.; Youngblood, N.; Karpov, M.; Gehring, H.; Li, X.; Gallo, M. L.; Fu, X.; Lukashchuk, A.; Raja, A.; Liu, J.; Wright, D.; Sebastian, A.; Kippenberg, T.; Pernice, W.; Bhaskaran, H. Parallel Convolution Processing Using an Integrated Photonic Tensor Core. *ArXiv200200281 Cond-Mat Physicsphysics* **2020**.

(20) Cheng, Z.; Ríos, C.; Pernice, W. H. P.; Wright, C. D.; Bhaskaran, H. On-Chip Photonic Synapse. *Sci. Adv.* **2017**, *3* (9), e1700160. https://doi.org/10.1126/sciadv.1700160.

(21) Zheng, J.; Khanolkar, A.; Xu, P.; Colburn, S.; Deshmukh, S.; Myers, J.; Frantz, J.; Pop, E.; Hendrickson, J.; Doylend, J.; Boechler, N.; Majumdar, A. GST-on-Silicon Hybrid Nanophotonic Integrated Circuits: A Non-Volatile Quasi-Continuously Reprogrammable Platform. *Opt. Mater. Express* **2018**, *8* (6), 1551–1561. https://doi.org/10.1364/OME.8.001551.

(22) Xu, P.; Zheng, J.; Doylend, J. K.; Majumdar, A. Low-Loss and Broadband Nonvolatile Phase-Change Directional Coupler Switches. *ACS Photonics* **2019**, *6* (2), 553–557. https://doi.org/10.1021/acsphotonics.8b01628.

(23) Tian, J.; Luo, H.; Yang, Y.; Ding, F.; Qu, Y.; Zhao, D.; Qiu, M.; Bozhevolnyi, S. I. Active Control of Anapole States by Structuring the Phase-Change Alloy Ge2Sb2Te5. *Nat. Commun.* **2019**, *10* (1), 396. https://doi.org/10.1038/s41467-018-08057-1.

(24) Wuttig, M.; Yamada, N. Phase-Change Materials for Rewriteable Data Storage. *Nat. Mater.* **2007**, *6* (11), 824–832. https://doi.org/10.1038/nmat2009.

(25) Xiong, F.; Liao, A. D.; Estrada, D.; Pop, E. Low-Power Switching of Phase-Change Materials with Carbon Nanotube Electrodes. *Science* **2011**, *332* (6029), 568–570. https://doi.org/10.1126/science.1201938.

(26) Li, X.; Youngblood, N.; Ríos, C.; Cheng, Z.; Wright, C. D.; Pernice, W. H.; Bhaskaran, H. Fast and Reliable Storage Using a 5 Bit, Nonvolatile Photonic Memory Cell. *Optica* **2019**, *6* (1), 1–6. https://doi.org/10.1364/OPTICA.6.000001.

(27) Cheng, Z.; Ríos, C.; Youngblood, N.; Wright, C. D.; Pernice, W. H. P.; Bhaskaran, H. Device-Level Photonic Memories and Logic Applications Using Phase-Change Materials. *Adv. Mater.* **2018**, *30* (32), 1802435. https://doi.org/10.1002/adma.201802435.

(28) Michel, A.-K. U.; Heßler, A.; Meyer, S.; Pries, J.; Yu, Y.; Kalix, T.; Lewin, M.; Hanss, J.; Rose, A. D.; Maß, T. W. W.; Wuttig, M.; Chigrin, D. N.; Taubner, T. Advanced Optical Programming of Individual Meta-Atoms Beyond the Effective Medium Approach. *Adv. Mater.* **2019**, *31* (29), 1901033. https://doi.org/10.1002/adma.201901033.

(29) Rios, C.; Stegmaier, M.; Cheng, Z.; Youngblood, N.; Wright, C. D.; Pernice, W. H. P.; Bhaskaran, H. Controlled Switching of Phase-Change Materials by Evanescent-Field Coupling in Integrated Photonics [Invited]. *Opt. Mater. Express* **2018**, *8* (9), 2455–2470. https://doi.org/10.1364/OME.8.002455.

(30) Zhang, W.; Mazzarello, R.; Wuttig, M.; Ma, E. Designing Crystallization in Phase-Change Materials for Universal Memory and Neuro-Inspired Computing. *Nat. Rev. Mater.* **2019**, *4* (3), 150–168. https://doi.org/10.1038/s41578-018-0076-x.

(31) Siegrist, T.; Jost, P.; Volker, H.; Woda, M.; Merkelbach, P.; Schlockermann, C.; Wuttig, M. Disorder-Induced Localization in Crystalline Phase-Change Materials. *Nat. Mater.* **2011**, *10* (3), 202–208. https://doi.org/10.1038/nmat2934.

(32) Ríos, C.; Zhang, Y.; Deckoff-Jones, S.; Li, H.; Chou, J. B.; Wang, H.; Shalaginov, M.; Roberts, C.; Gonçalves, C.; Liberman, V.; Gu, T.; Kong, J.; Richardson, K.; Hu, J. Reversible Switching of Optical Phase Change Materials Using Graphene Microheaters. In *2019 Conference on Lasers and Electro-Optics (CLEO)*; 2019; pp 1–2. https://doi.org/10.1364/CLEO_SI.2019.SF2H.4.

(33) Singh, V.; Joung, D.; Zhai, L.; Das, S.; Khondaker, S. I.; Seal, S. Graphene Based Materials: Past, Present and Future. *Prog. Mater. Sci.* **2011**, *56* (8), 1178–1271. https://doi.org/10.1016/j.pmatsci.2011.03.003.

(34) Zheng, J.; Fang, Z.; Wu, C.; Zhu, S.; Xu, P.; Doylend, J. K.; Deshmukh, S.; Pop, E.; Dunham, S.; Li, M.; Majumdar, A. Nonvolatile Electrically Reconfigurable Integrated Photonic Switch. *ArXiv191207680 Phys.* **2019**.

(35) Zhang, H.; Zhou, L.; Lu, L.; Xu, J.; Wang, N.; Hu, H.; Rahman, B. M. A.; Zhou, Z.; Chen, J. Miniature Multilevel Optical Memristive Switch Using Phase Change Material. *ACS Photonics* **2019**, *6* (9), 2205–2212. https://doi.org/10.1021/acsphotonics.9b00819.

(36) Zhang, H.; Zhou, L.; Xu, J.; Wang, N.; Hu, H.; Lu, L.; Rahman, B. M. A.; Chen, J. Nonvolatile Waveguide Transmission Tuning with Electrically-Driven Ultra-Small GST Phase-Change Material. *Sci. Bull.* **2019**, *64* (11), 782–789. https://doi.org/10.1016/j.scib.2019.04.035.

(37) Weijtens, C. H. L. Influence of the Deposition and Anneal Temperature on the Electrical Properties of Indium Tin Oxide. *J. Electrochem. Soc.* **1991**, *138* (11), 3432–3434. https://doi.org/10.1149/1.2085429.

(38) Ma, Z.; Li, Z.; Liu, K.; Ye, C.; Sorger, V. J. Indium-Tin-Oxide for High-Performance Electro-Optic Modulation. *Nanophotonics* **2015**, *4* (1), 198–213. https://doi.org/10.1515/nanoph-2015-0006.

(39) Song, S.; Yang, T.; Liu, J.; Xin, Y.; Li, Y.; Han, S. Rapid Thermal Annealing of ITO Films. *Appl. Surf. Sci.* **2011**, *257* (16), 7061–7064. https://doi.org/10.1016/j.apsusc.2011.03.009.

(40) Aspnes, D. E. Local-field Effects and Effective-medium Theory: A Microscopic Perspective. *Am. J. Phys.* **1982**, *50* (8), 704–709. https://doi.org/10.1119/1.12734.

**Figures and Figures' Captions:**

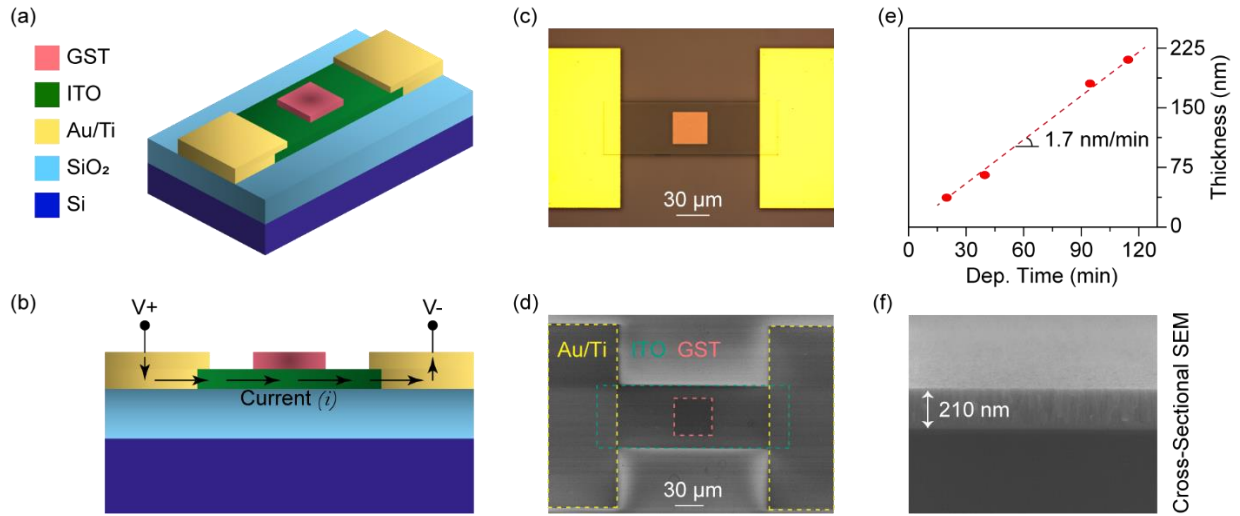

**Figure 1: ITO microheaters.** The schematic illustration of designed microheaters depicted in (a) top and (b) side views. The Joule heating in the ITO layer serves as the heat source for the phase conversion of the integrated GST patch. (c) Optical and (d) SEM images of fabricated devices. (e) Thickness of the GST layer as a function of the sputtering time (RF power: 40 W, vacuum: $10^{-2}$ Torr). A linear deposition rate of ~ 1.7 nm/min is obtained. (f) A representative cross-sectional SEM image that shows the thickness uniformity of deposited GST films over micrometer scales.

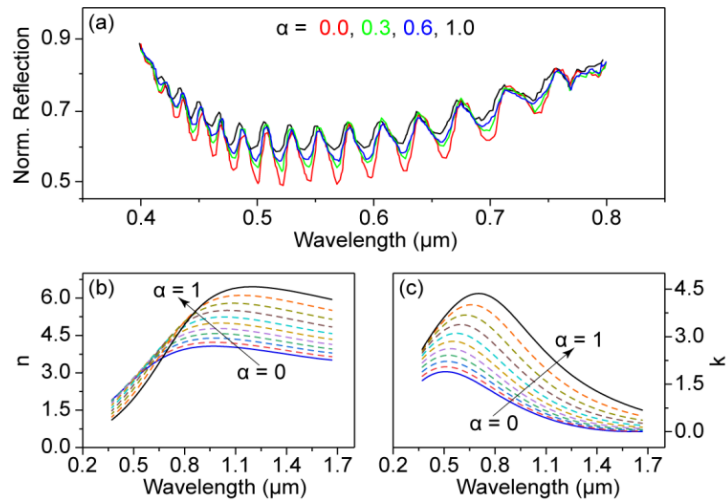

**Figure 2: Characterization of the GST phase change using optical reflection spectroscopy.** (a) Optical reflection spectra of the GST/ITO/SiO$_2$/Si stack normalized to that of the ITO/SiO$_2$/Si stack. The crystallization degree (i.e., α) after each Joule heating event is extracted using the transfer-matrix method (see SI for details). As explained in the main text, an effective-medium approximation is assumed for the optical properties of the GST layer. (b, c) Refractive index (n) and extinction coefficient (k) of GST, respectively, at fully amorphous (i.e., α = 0), fully crystalline (i.e., α = 1), and nine intermediate states with α = 0.1 incremental steps. Solid lines are experimentally measured by spectroscopic ellipsometry. Dashed lines are extrapolated using the Lorentz-Lorenz equation as explained in the main text.

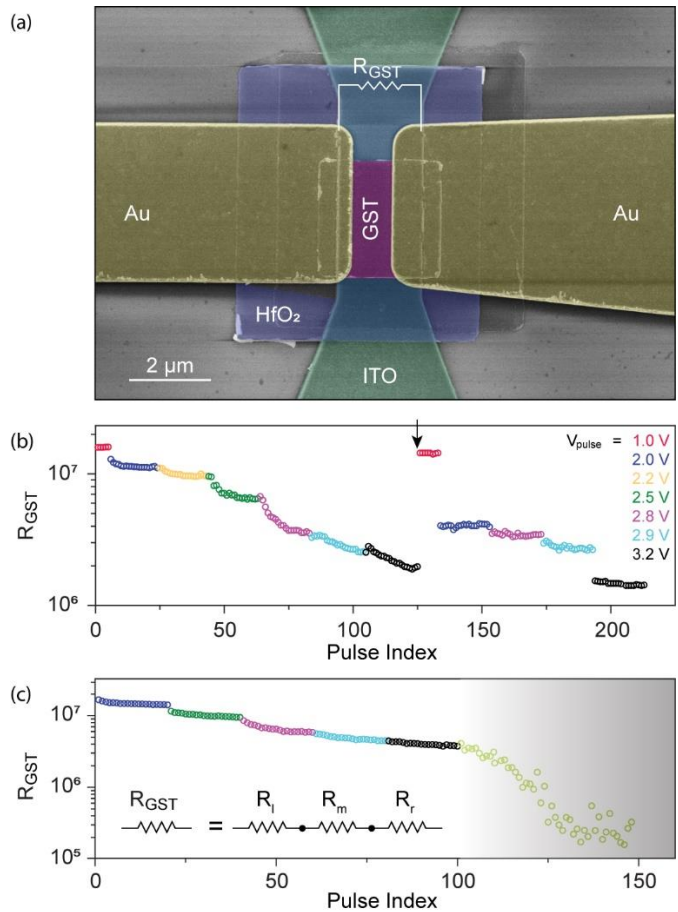

**Figure 3: Multi-stage and reversible switching of the GST phase.** (a) False-colored SEM image of the fabricated device. The GST phase is changed through the Joule heating by the underlying ITO layer and monitored by the measurement of $R_{GST}$ between the Au contacts. (b) The GST resistance as a function of the number and amplitude of voltage pulses applied to the ITO heater. All pulses are 200 msec wide; their amplitudes range from 1 V to 3.2 V. The arrow highlights the re-amorphization of GST by the application of a short voltage pulse (width: 50 ns, amplitude: 9 V) to the ITO microheater. (c) Role of different effects in the variation of $R_{GST}$ with the crystalline phase of the GST patch in panel (a). Inset: the measured $R_{GST}$ is composed of two side segments close to the left and right electrodes ($R_l$ and $R_r$) and a middle segment ($R_m$) in panel (a). As explained in the text, Joule heating in ITO (i.e., the unshaded part of panel (c), for pulse index < 100) primarily affects $R_m$. The side segments (i.e., $R_l$ and $R_r$) can be affected by the application of current pulses to the Au electrodes as shown in the shaded region (beyond pulse index 100). The width and amplitude of current pulses applied to the two Au electrodes are 200 msec and 200 µA, respectively.

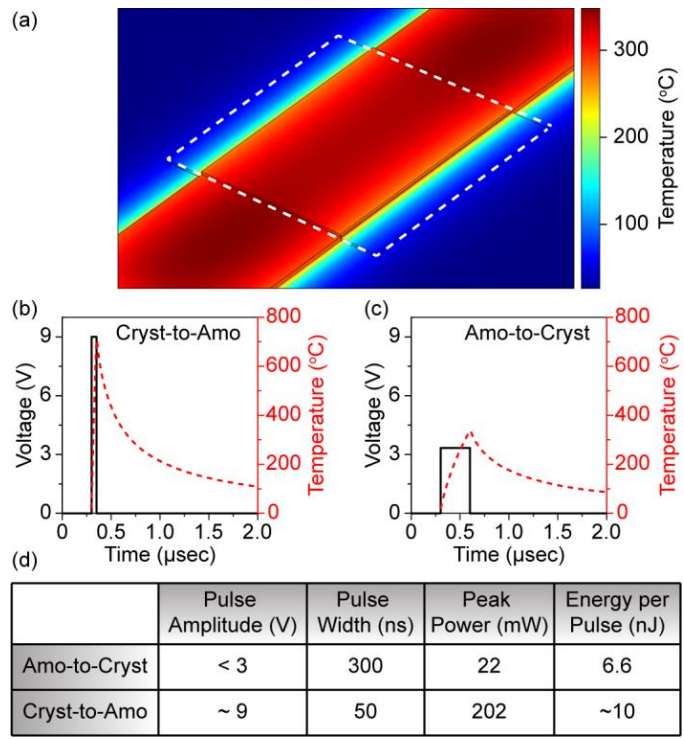

**Figure 4: Electro-thermal simulation of the ITO microheater.** (a) The spatial heat profile across the hybrid ITO/GST device in Figure 3(a) plotted at the end of a 300 nsec-wide electric pulse with a 3V amplitude. Dashed lines outline the GST patch. Geometrical dimensions are similar to those shown in Figure 3(a) without the readout electrodes. (b, c) Temporal heat profiles (dashed lines) generated by the voltage pulses (solid lines) applied to the ITO microheater for amorphous-to-crystalline and crystalline-to-amorphous phase switching, respectively. (d) A summary of the pulse condition and power/energy consumption for the reversible switching of the GST phase. For these calculations, the ITO resistance was measured to be 400 Ω.

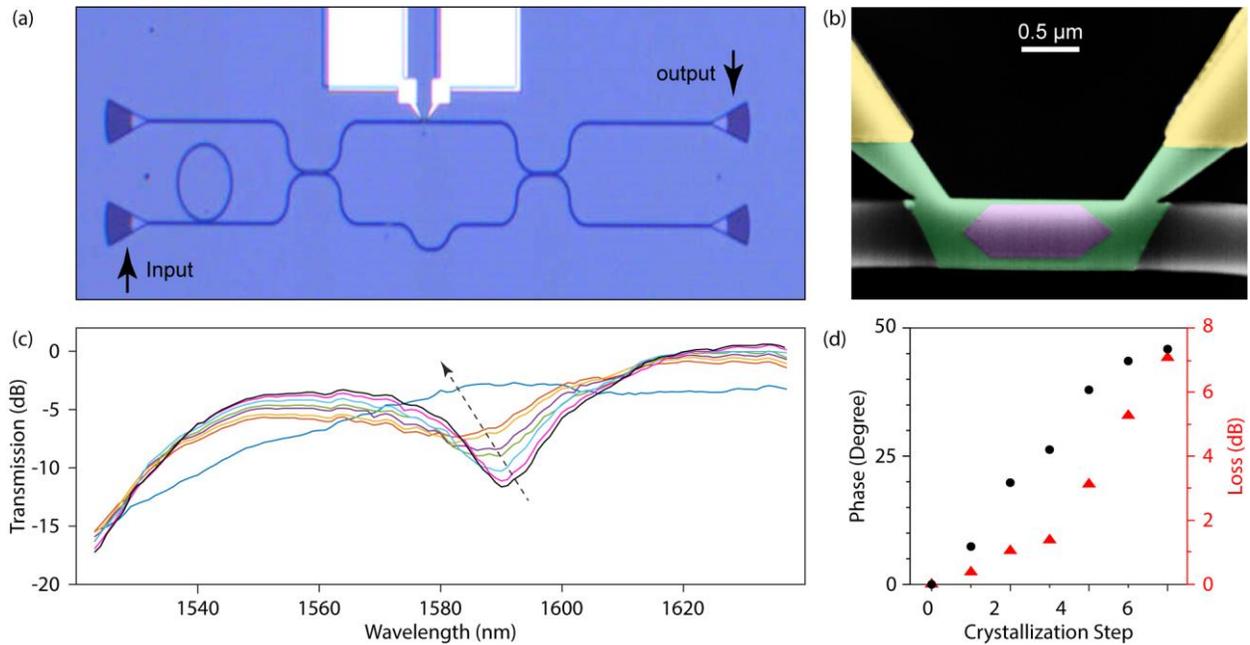

**Figure 5. Multi-stage MZI phase-shifter.** (a) Optical image of a MZI phase shifter fabricated in a 400 nm-thick SiN platform. Input and output ports are identified on the image. The ring resonator is used for the calibration purpose to compensate for potential environmental changes. (b) The false-colored SEM image of the GST/ITO segment integrated on the upper arm of the MZI in (a). GST, ITO, and Au layers are colored purple, green, and yellow, respectively. (c) The optical transmission (i.e., from input to output) for several crystalline states of GST. The arrow points from the amorphous state to the crystalline state. (d) Extracted phase shifts and optical loss for each crystalline state of the GST from the experimental results in (b). (e) The FDTD simulation of the electric field profile along the SiN waveguide for amorphous (top) and crystalline states of the integrated GST section. The field overlap with the GST section increases following the crystallization of GST. Field profiles are separately normalized for amorphous and crystalline states.